# Earth-Affecting Solar Causes Observatory (EASCO): A mission at the Sun-Earth L5

Nat Gopalswamy*[a], Joseph M. Davila[a], Frédéric Auchère[b], Jesper Schou[c], Clarence Korendike[d], Albert Shih[a], Janet C. Johnston[e], Robert J. MacDowall[a], Milan Maksimovic[f], Edward Sittler[a], Adam Szabo[a], Richard Wesenberg[a], Suzanne Vennerstrom[g], and Bernd Heber[h]

[a]NASA Goddard Space Flight Center, Greenbelt, MD, USA 20771; [b]Institut d'Astrophysique Spatiale, CNRS/Université Paris-Sud 11, Bâtiment 121, 91405 Orsay, France, [c]Center for Space Science and Astrophysics, Stanford University, Stanford, CA USA 94305-4085, [d]Naval Research Laboratory, 4555 Overlook Ave., SW Washington, DC USA 20375, [e]Air Force Research Laboratory, Space Vehicles Directorate 29 Randolph Road, Hanscom AFB, MA USA 01731-3010, [f]LESIA,University of Paris Meudon, [g]National Space Institute, Technical University of Denmark, Juliane Maries vej 30, 2100 København Ø Denmark, [h]Christian Albrecht University, Leibnizstr. 11, 24118 Kiel, Germany

## ABSTRACT

Coronal mass ejections (CMEs) and corotating interaction regions (CIRs) as well as their source regions are important because of their space weather consequences. The current understanding of CMEs primarily comes from the Solar and Heliospheric Observatory (SOHO) and the Solar Terrestrial Relations Observatory (STEREO) missions, but these missions lacked some key measurements: STEREO did not have a magnetograph; SOHO did not have in-situ magnetometer. SOHO and other imagers such as the Solar Mass Ejection Imager (SMEI) located on the Sun-Earth line are also not well-suited to measure Earth-directed CMEs. The Earth-Affecting Solar Causes Observatory (EASCO) is a proposed mission to be located at the Sun-Earth L5 that overcomes these deficiencies. The mission concept was recently studied at the Mission Design Laboratory (MDL), NASA Goddard Space Flight Center, to see how the mission can be implemented. The study found that the scientific payload (seven remote-sensing and three in-situ instruments) can be readily accommodated and can be launched using an intermediate size vehicle; a hybrid propulsion system consisting of a Xenon ion thruster and hydrazine has been found to be adequate to place the payload at L5. Following a 2-year transfer time, a 4-year operation is considered around the next solar maximum in 2025.

**Keywords:** Coronal mass ejections, corotating interaction regions, Sun-Earth L5, solar electric propulsion, helioseismology, coronagraph, heliospheric imager, EASCO mission design

## 1. INTRODUCTION

Only after the advent of the Solar and Heliospheric Observatory (SOHO) mission has it been recognized that CMEs represent the most energetic phenomenon in the Heliosphere. CMEs derive their energy from the stressed magnetic fields on the Sun that are concentrated in solar active regions. Initiation of CMEs and their early evolution are poorly understood. CMEs affect Earth in two ways: (i) by accelerating high energy protons and electrons that pose serious radiation hazard to space systems and astronauts. (ii) CMEs also produce the largest of geomagnetic storms that can severely impact the society by disrupting infrastructure such as power grids, pipelines, and railroads. CIRs cause moderate storms, during which the magnetospheric electrons are accelerated to MeV range and hence pose a radiation hazard to satellites operating within the magnetosphere. The wealth of knowledge on CMEs accumulated over the last three decades has been from coronagraphs located along the Sun-Earth line (ground-based or space-borne). Coronagraphs by their very nature cannot directly observe CMEs heading towards Earth because of the occulting disk, which blocks that part of the Earth-directed CMEs that arrives at Earth. It is likely that the occulting disk also blocks the nose of the CME-driven shock, where the shock is strongest and likely to accelerate particles. The case of CIRs is worst because we can observe them only when they are about to hit Earth (except for the white-light observations using Heliospheric Imager on board STEREO). Therefore, one needs to observe CMEs and CIRs from a different vantage point that provides a full view of CMEs when they are still close to the Sun and CIRs well before they arrive at Earth. The Lagrange point L5 is ideally suited for remote sensing CMEs and observing CIRs in-situ.

*nat.gopalswamy@nasa.gov; phone 1 301-286-885; fax 1 301-286-7194

It must be noted that SOHO and STEREO missions did not carry all the instruments ideally suited to study the complete evolution of CMEs and CIRs. For example, SOHO did not carry a magnetometer for in-situ observations or a radio instrument for remote sensing CME-driven shocks. STEREO did not carry a magnetograph to observe the magnetic properties of the CME and CIR source regions. A new mission located at Sun-Earth L5 can solve these problems and provide an unprecedented wealth of data on CMEs and CIRs[1-3].

The Earth Affecting Solar Causes Observatory (EASCO) mission concept has already been described in details, especially the science issues and the measurements that need to be made to tackle these issues[3]. EASCO was recently studied at the Mission Design Laboratory (MDL) of NASA's Goddard Space Flight Center. This paper reports the results of this MDL study.

## 2. OVERVIEW OF THE EASCO MISSION

### 2.1 Science Requirements

The key science requirements of the EASCO mission are as follows:

(i) Observe CMEs, CIRs, and shocks from a vantage point (Sun-Earth Lagrange point L5) that avoids projection effects thus providing a full view of each particular event, Sun to Earth. (ii) Observe the magnetic and velocity fields at the photospheric layer to study the origin of solar magnetism and its consequences in the solar atmosphere in the form of sunspots, active regions, and coronal holes. (iii) Observe the solar source region and processes that are associated with e these large-scale solar events. (iv) Observe continuously each event, with a range of in-situ and remote-sensing instruments, as it propagates from Sun to Earth (CMEs and the shocks driven by them travel from Sun to Earth from < 1 to about 4 days). (v) Make in-situ measurements of the plasma and magnetic field content of CIRs that travel from L5 to Earth in ~4 days.

### 2.2 Scientific Instruments

The scientific payload of the EASCO mission that meets the science requirements listed in section 2.1 consists of seven remote-sensing and three in-situ instruments. The remote-sensing instruments observe signatures of magnetism in the solar interior and its various manifestations in the solar atmosphere that ultimately result in the large-scale disturbances such as CMEs and CIRs.

The Magnetic and Doppler Imager (MADI) will measure the photospheric magnetic and velocity fields, map the photospheric magnetic field and study the physical conditions in the tachocline. The Doppler images from MADI can be combined with those obtained by other instruments such as HMI and the Global Oscillations Network (GONG) for probing various subsurface layers. Magnetograms obtained from L5 view can also track the complexity and evolution of the active regions and coronal holes well before they come to the Earth view. Surface magnetic field measurements from L5 will be helpful to modeling efforts that depend on extrapolation of the photospheric field. MADI will be a scaled-down version of the Helio Magnetic Imager (HMI) on board the Solar Dynamics Observatory (SDO) and SOHO/MDI with additional modifications to reduce the resources (mass power, and data rate).

The Inner Coronal Imager in EUV (ICIE) will provide advance warning of active regions and coronal holes soon to be rotating on to the Earth view, complementary to the information provided by MDI at the photospheric level. ICIE is crucial in identifying the source regions of CMEs and can provide information on the early evolution of CMEs enabling the determination of the initial acceleration[4]. ICIE can also greatly improve the 3-D modeling of EUV irradiance using Carrington maps[5] by providing information of the corona behind the east limb thereby reducing the error in the maps. Imagers at EUV wavelengths have attained a high level of maturity and the ongoing research and development activities greatly help the design of ICIE. At least two wavelengths are needed (one corresponding to the cold and hot plasmas), but more wavelengths can be used if resources permit. ICIE will have a 4 solar radii (Rs) field of view (FOV).

The white-light coronagraph (WCOR) will image CMEs near the Sun in the heliocentric distance range of 1.5 – 15 Rs. Measuring CMEs in this distance range is crucial for propagation models that predict the Earth-arrival time of CMEs. Observing CMEs close to the Sun provides advanced warning by days. The inner edge of the WCOR FOV will be close to the outer edge of the ICIE FOV for tracking CMEs from close to the Sun. WCOR is an externally occulted, refracting, Lyot visible light coronagraph, with a polarization analyzer, providing spatial resolution in the range 5-10 arcsec.

The Heliospheric Imager (HI) is a very-wide, white-light coronagraph with a FOV in the range 15° to 150° from the Sun center. CME features from the outer edge of the WCOR FOV can be tracked into the HI FOV and hence all the way to Earth. CME tracking can help validate CME propagation models that take WCOR measurements as input. HI enables measurement of crucial, fine-scale features near or on the spacecraft orbit, in addition to the global structure of the corona.

The Hard X-ray Imager (HXI) will image the non-thermal component of flares. The HXI images provide direct information on the height of the energy release in the source of Earth-directed CMEs. Combining HXI, ICIE, and WCOR images, one can study the flare-CME relationship and the two sites of particle acceleration (flare reconnection and CME-driven shock). The proposed HXI is based on the Solar Orbiter design. Measured visibilities (Fourier components of the image) are combined on the ground to reconstruct images. The spatial resolution is in the range 4 – 100 arcsec; the spectral resolution is in the range 1 – 15 keV in the operating energy range of 4 – 150 keV; the time resolution is 0.1s.

Table 1. Mass, power, and telemetry requirements for EASCO science instruments

| Instrument | Mass (kg) | Power (W) | Data Rate (kbps) |
|---|---|---|---|
| MADI | 15 | 60 | 7 |
| ICIE | 10 | 8 | 30 |
| WCOR | 25 | 30 | 15 |
| HI | 10 | 15 | 5 |
| HXI | 6 | 5 | 2 |
| UVOS | 30 | 30 | 20 |
| LRT | 13 | 15 | 2 |
| SWPI | 10 | 5 | 3 |
| MAG | 3 | 3 | 3 |
| EPD | 16 | 23 | 3 |
| Total | 138 | 194 | 90 |

The Ultra-Violet Off-limb Spectrograph (UVOS) will measure shock properties of Earth-directed CMEs at two coronal heights (2 and 3.5 Rs). Solar energetic particles are thought to be released at these heights, so continuous monitoring of the corona will provide information on the physical conditions in the ambient medium and seed particles that get accelerated by the CME-driven shocks. The spectrograph operating in the wavelength range 970-1040 (first order) and 435-520Å (second order) with continuous monitoring at two heights using a 3-Rs long slit is readily achievable with current state of the art. Several bright lines exist in the proposed wavelength range that can be used for diagnostics with a spectral resolution of 0.1 Å and a spatial resolution of ~20 arcsec.

The Low-frequency Radio Telescope (LRT) will isolate CMEs driving shocks near the Sun using type II radio bursts in the decameter-hectometric (DH) wavelength range. The drift rate of type II bursts provide the shock speeds that can be compared with the shock and CME speeds measured from white light. For Earth-directed events, the shock properties from UVOS can be calibrated against the LRT observations. LRT will obtain three-axis radio spectrum from the local plasma frequency at the spacecraft to about 15 MHz, very similar to the STEREO/WAVES Experiment.

The Solar Wind Plasma Instrument (SWPI) will make in-situ measurement of the solar wind, so the CIRs can be characterized ~4 days ahead of Earth arrival. SWPI consists of a Fast Electron Analyzer (FEA) and an Ion Composition Analyzer (ICA). FEA measures solar wind electrons over a wide angular range, including the nadir direction (to provide direct measure of strahl electrons). ICA measures the solar wind ions, pick-up ions, and low energy suprathermal ions providing mass, composition and charge states over nearly 360° in azimuth and scanning ±45° in elevation.

The solar wind magnetometer (MAG) will measure the magnetic content of CIRs ~4 days before Earth arrival. MAG is a single, triaxial, wide-range, low power, low noise fluxgate magnetometer mounted on a 3-m boom, obtaining 32 samples per second.

The Energetic Particle Detector (EPD) will detect energetic particles from solar eruptions that will help understand flare and CME-shock contributions to large SEP events. EPD consists of a Low Energy Telescope (LET) to measure ions including protons in the energy range 1.5 to 12 MeV and a High Energy Telescope (HET) to measure ions (from 20 to 200 MeV/nucleon), electrons (~0.3 to 20 MeV), and gamma rays (0.3-10 MeV). HET also measures ion composition (He–Fe and $^3$He/$^4$He).

Table 1 shows the resource requirements of the ten science instruments. The EASCO mission needs to carry a total of ~138 kg of payload mass to L5 and 194 W of power to operate the instruments that will send data to ground stations at the rate of 90 kbps.

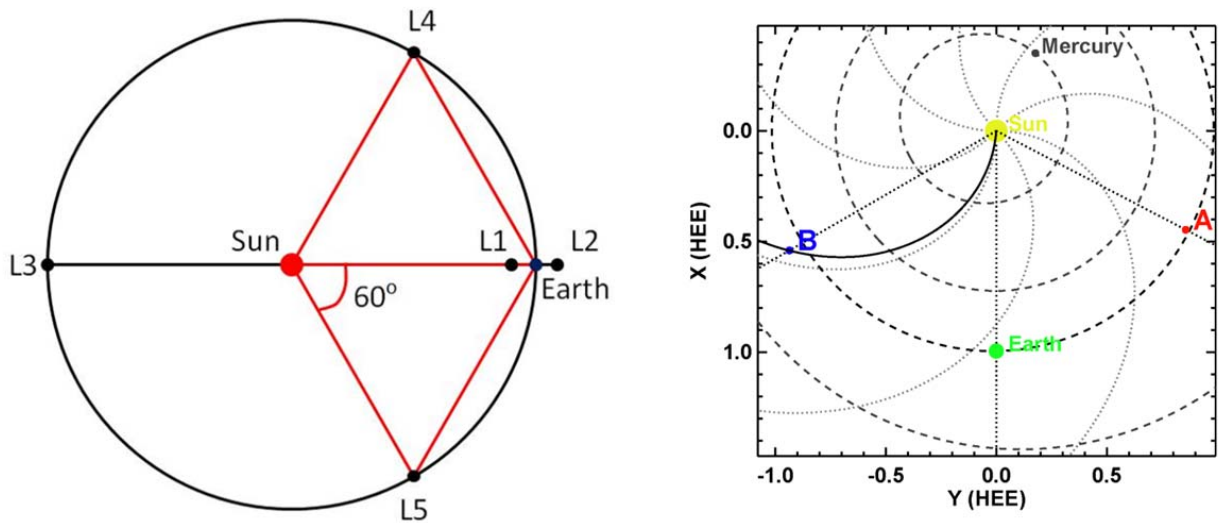

Figure 1. (left) The five Lagrange points of the Sun-Earth gravitational system. The EASCO mission is proposed to be located at L5. (right) The relative positions of the STEREO Ahead (A) and Behind (B) spacecraft with respect to Earth on 25 October 2009 when B was near L5 and A was near L4. The thick spiral starting from the Sun and passing near B represents a CIR, which would rotate to the position of Earth in ~4 days.

### 2.3 Observation Geometry

The five Lagrange points of the Sun-Earth gravitational system are shown in Fig. 1. The EASCO mission will be located at L5. Note that SOHO has been observing from L1, which is located along the Sun-Earth line. CMEs heading toward Earth can be observed from L5 in a broad-side view, so measurements can be made without projection effects. Figure 1 (right) shows the locations the STEREO Ahead (A) and Behind (B) spacecraft on 25 October 2009 when they were located close to the L4 and L5 points, respectively. In the case of EASCO, the spacecraft will be stationed at L5 and such that the final spacecraft speed matches that of Earth around the Sun. The spacecraft will have the same eccentricity as Earth. Note that a spacecraft located at Sun-earth L4 will be equally suitable for measuring Earth-directed CMEs, but from L4 one cannot observe new active regions emerging/rotating behind the east limb in the Earth view. Furthermore, CMEs producing solar energetic particle (SEP) events at Earth generally head toward L4, so they cannot be measured without projection effects from L4. From L5, one can measure the shock structure ahead of CMEs and it can be used to derive the heliospheric magnetic field in the upstream medium[6].

### 2.4 Target Performance Period

The EASCO mission will commence operations preferentially one year before the maximum of solar cycle 25, which is expected to occur around the year 2025. EASCO will observe on-station from L5 for 4 years. The goal is to continue the operations until the following maximum that will occur ~11 years later. During the transfer phase between Earth and L5, additional science observations are possible, but only by the instruments not affected by the maneuvers. The transfer phase is expected last ~2 years. Thus the target launch of the EASCO mission was assumed to be at beginning of the year 2022 for the MDL run.

### 2.5 Operating Modes

The primary operating mode is scientific observing. Some instruments need calibration mode. EASCO will provide the means to support space weather beacon mode; providing a timely useful set of space environment conditions to evaluate the feasibility of forecasting the variant space weather with the goal to develop early warning schemes.

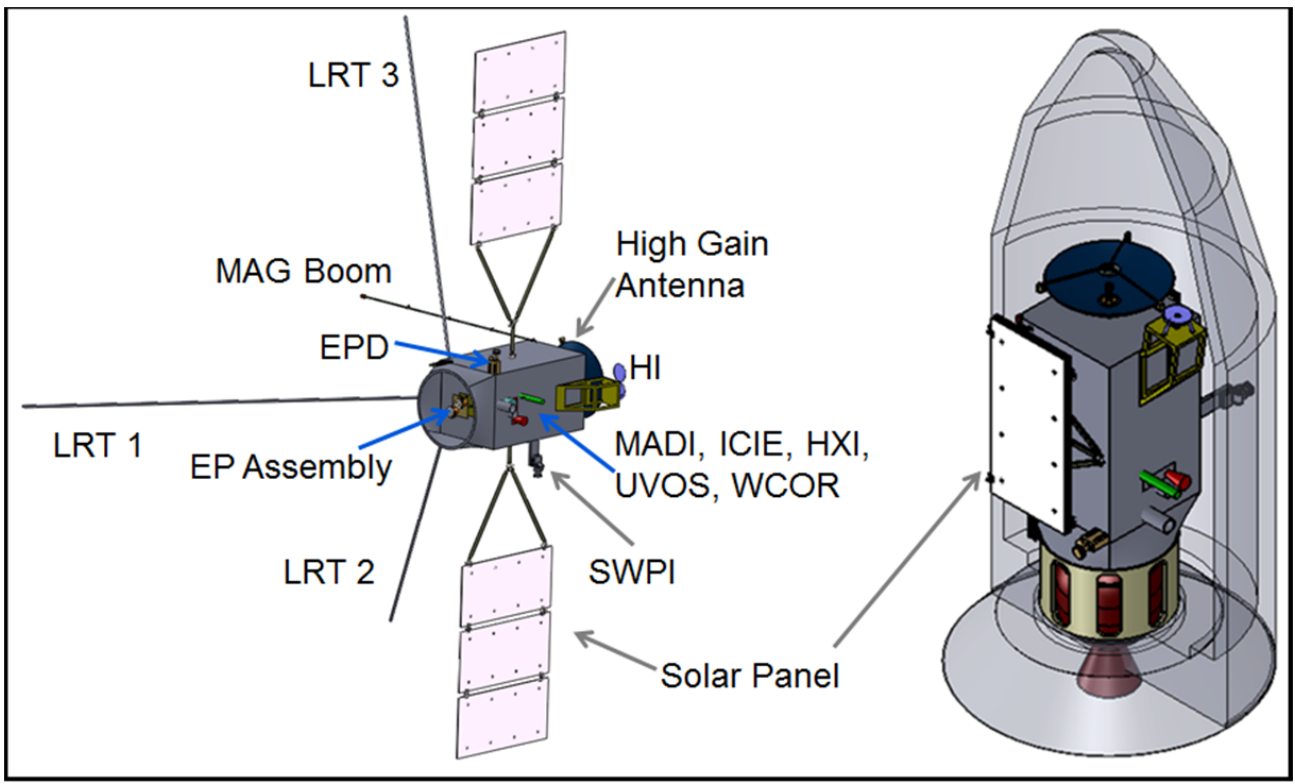

Figure 2. (left) The EASCO spacecraft and the accommodation of the ten scientific instruments. (right) The launch configuration of the EASCO mission using a launch vehicle similar to Taurus II with an enhanced faring and third stage.

### 3. RESULTS OF THE MDL STUDY

The MDL study was conducted by the EASCO team in collaboration with the GSFC engineers during March 7 – 11, 2011. The study focused on all the four segments of the EASCO system architecture: Observatory (science payload, spacecraft, and ground stations), Launch (vehicle and launch site services), Mission Operations and Data Analysis (MO&DA including Mission Operations Center, Science Operations Center, and Communications), Facilities, Test and Handling. Particular attention was paid to the flight dynamics considering both chemical options and electric propulsion.

### 3.1 Mechanical Assembly

The spacecraft bus with the ten scientific instruments, the high-gain antenna (HGA), the solar panels, the electric propulsion (EP) assembly are shown in Fig. 2 along with the launch configuration. The S/C bus will be a rectangular composite honeycomb structure, with a 62" separation system. The cluster of remote sensing telescopes (MADI, ICIE,

HXI, UVOS, and WCOR) is placed together on the Sun-facing side of the spacecraft. The HI instrument is mounted on a platform to clear the HGA. All subsystem components, instruments, and instrument electronics can be accommodated within the spacecraft bus envelope. The bus shown is not densely packed, allowing for potential observatory growth or scaling down of the structure. The spacecraft bus structural design and fabrication is standard practice, which includes: Machined aluminum alloy structural members – corner posts, propulsion structure, radials, secondary structure, mounts for solar arrays and the HGA; composite skin with aluminum alloy core honeycomb panels, decks, launch vehicle interface to bus structure interface adapter. The launch assembly shown in Fig. 2 can be readily accommodated in a launch vehicle similar to Taurus II with enhanced faring and a third stage. The HGA is at the top, so it will be facing away from Earth in the beginning of the cruise phase. The spacecraft will be rolled midcourse such that the HGA points to Earth for the prime mission telemetry.

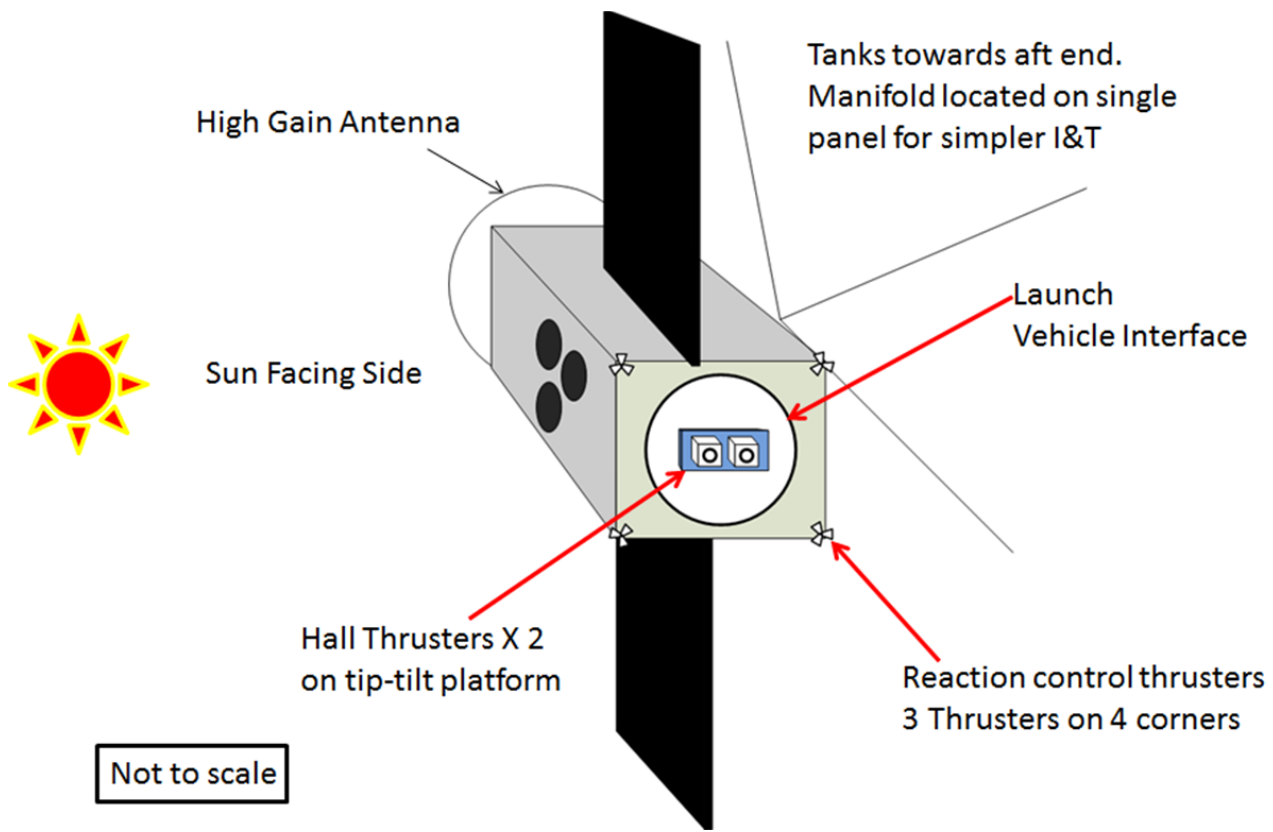

Figure 3. The EASCO electric propulsion system, located on the anti-Earth side of the spacecraft. While the Hall thrusters are mounted on a tip-tilt platform, the reaction control thrusters are mounted on the four corners of the S/C bus.

### 3.2 Propulsion

One of the important outcomes of the MDL study is that a hybrid propulsion system that combines solar Electric Propulsion (EP) and monopropellant hydrazine (MH) is the best option to place the EASCO mission at L5. This was determined after a trade study was conducted of different architectures. The EP-MH hybrid architecture provides lower mass for the mission and greater flexibility on launch vehicles. Purely-chemical propulsion would require six times more fuel mass. The hybrid system also provides flexibility in the choice of launch vehicle and puts the observatory within Taurus II's capability. EP system will be used for the large delta V maneuvers and trajectory correction, using two Hall thrusters (primary and secondary), a power processing unit that controls the thrusters and xenon feed. Thrusters will be articulated to align thrust vector with spacecraft center-of-mass. The thrusters will be located on the anti-earth face of the spacecraft (see Fig. 3). The system will have a throttling ability that will allow taking the best advantage of power available. A simple Mono-propellant Blow-Down system (specified for operating pressure of 350 psia) will be used to unload momentum. This will require about 8 kg of ultra-pure hydrazine fuel. EP has been used in many commercial satellites, but only in a few scientific missions: European Space Agency's SMART-1 mission[7] (Small Missions for Advanced Research in Technology) and NASA's DAWN mission[8]. SMART-1 used a solar-powered Hall-effect thruster with xenon propellant, weighing 82 kg at launch. The DAWN mission also uses hybrid propulsion and carries a 45 kg hydrazine tank and a 450 kg xenon tank. Japan's Hayabusa mission also employed ion thrusters[9].

### 3.3 Flight Dynamics

The flight dynamics study focused on finding the best orbit to get to L5 and stop the spacecraft and make it attain Earth's speed around the Sun with similar eccentricity. It was also required (for MADI) that the radial velocity of the spacecraft at the final location does not exceed that of Earth (0.5 km/s). Final orbit is targeted by directing MALTO (Mission Analysis for Low-Thrust Optimization) software to fly to a rendezvous with a fictitious body in Earth's orbit and trailing Earth by 60° (i.e., L5). A final orbit with a semi-major axis of 1 AU and an eccentricity of 0.0167 or less meets the radial velocity constraint. The baseline low-thrust trajectory to Sun-Earth L5 using EP-MH hybrid thruster takes a transfer time of ~2 years for a launch C3 (Earth-escape energy) ~2.2 km$^2$/s$^2$ and a delta-V of ~1.5 km/s. The required propellant mass (xenon) is: ~55 kg. Optional high-thrust trajectories using chemical propulsion (CP) were also considered. For the same transfer time, the required launch C3 is ~1.0 km$^2$/s$^2$ and the delta-V is ~950 m/s. The major difference is the required propellant mass: ~300 kg in the case of CP.

### 3.4 Power

Since the EP system uses power from the solar panels, the array needs to have a total area of 13.71 m$^2$. The solar array is segmented onto two power buses: 3.63m$^2$ for the spacecraft and instruments and 10.08m$^2$ for the cruise phase low-thrust propulsion. A converter or regulator is provided to make power from the low-thrust propulsion bus available to the spacecraft and instruments after the cruise phase. An array operating temperature of 70°C and a Spenvis solar array degradation prediction Pmax 2.282E+14 at the fifteen year point is assumed. In addition, the solar array must be electrostatically clean with a way to bleed off impinging charges, must avoid electrostatic discharges, and be magnetically quiet. A battery is provided for launch and orbit insertion. After arrival at L5 the battery is available for contingencies. A 24 Ah ABSL battery was selected, which has multiple parallel strings. The power system electronics are integrated with the spacecraft avionics module.

### 3.5 Communications

An overview of the EASCO communications system is provided in Fig. 4. EASCO will use the Deep Space Network (DSN) for science dump, command, and ranging. The DSN ground station is a 34 m dish. For the space weather beacon (SWB) mode data, a 9-m NOAA ground station will be used. EASCO has three antennas: (i) the HGA, which is a 1.2 m, dual frequency (Ka and X bands), parabolic reflector used for nominal mission support. (ii) a medium gain antenna (MGA), which is a double helix used for early cruise phase, when the HGA is on the anti-Earth face of the spacecraft). (iii) Omni antennas will be located 180° apart with one being placed as close to Earth pointing as possible; they will support during launch and early orbit, or in an emergency. The MGA and the omni antennas will operate in X band.

The Ka-Band downlink to DSN 34 m at the rate of 363 kbps at max range using 100 W TWTA and 20° station elevation would take about 8.2 hours/day. If there is significant data compression, larger antenna, or more power, the downlink time can be reduced. The X-Band HGA downlink to DSN 34 m will be at the rate of 80 kbps at max range using 55 W TWTA. The beacon mode will provide near continuous space weather data (500 bps) via the HGA X-band to 9-m systems when on location. The X-band MGA support is nominally to 0.5 AU (before mid-cruise flip). The data rate id 825 bps at max range for all configurations. Higher rate (2.7 kbps) is available when in the center beam of the antenna with 0 dB margin. The near-Earth rate is greater than 80 kbps. For the X-band omni support, the data rates are also variable: 22 bps at max range for all configurations and > 80 kbps for near-Earth support.

EASCO commanding will be at the rate of 2 kbps. For the HGA, the 2 kbps rate is possible at all times (after mid-cruise flip). For the MGA, the 2 kbps rate is possible at all times (prior to 0.5 AU, i.e., before the mid-cruise flip). For the omni antennas, the 2 kbps is possible at near Earth but the rate decreases significantly at L5 (100 bps or 22 bps worst case).

Ranging will use X-band up and Ka-band down whenever ranging is required and the HGA is pointed to the Earth. During the first half of the cruise phase, ranging will be in X-band both up and down. Ranging will use X-band up and Ka-band down whenever ranging is required and the HGA is pointed to the Earth. Ranging can be accomplished using

X-band up and down with any antenna. Ranging can be simultaneous with telemetry but could reduce the Ka-band downlink data rate. The nominal scenario is to transmit science and housekeeping data via Ka-band while simultaneously ranging via X-band. Note that during the cruise phase, ranging and housekeeping will be via X-band because HGA is not Earth-pointed.

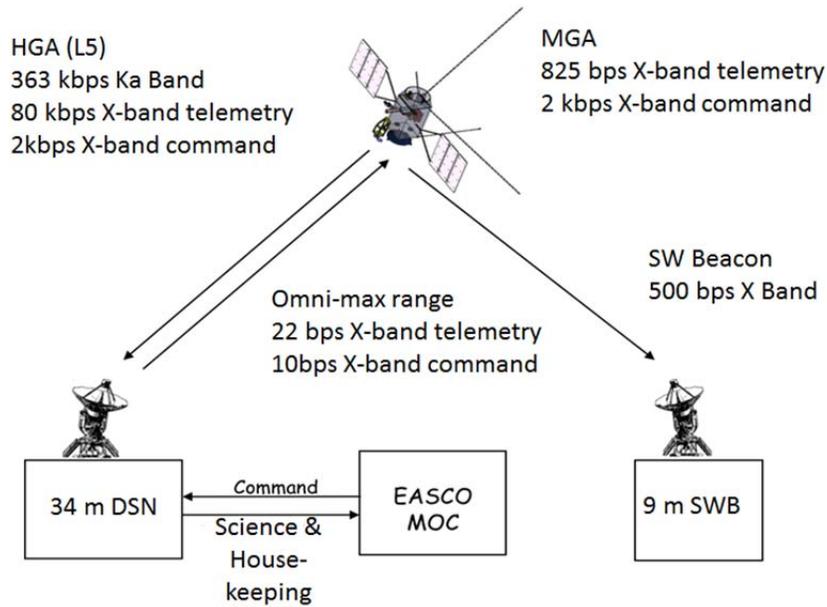

Figure 4. (left) Communications overview for the EASCO mission. DSN 34 m ground station will be used for science and house-keeping data. A 9 m ground station will be used for operation in the space weather beacon (SWB) mode. Three antenna systems will be used on the spacecraft: The high gain antenna (HGA), the medium gain antenna (MGA) and omni antennas. EASCO Mission Operations Center (MOC) receives science and house-keeping data and sends commands via the 34 m.

Table 2. Launch vehicle evaluation based on the EASCO system mass

| Spacecraft (S/C) Dry Mass | | | |
|---|---|---|---|
| | CBE | Contingency | Allocation |
| Payload Total | 138 kg | 29% | 178 kg |
| S/C Bus Total | 559 kg | | 657 kg |
| S/C Dry Mass | 698 kg | | 835 kg |
| Xenon | 55 kg | 0% | 55 kg |
| Hydrazine | 10 kg | 0% | 10 kg |
| S/C Wet Mass | | | 900 kg |
| Separation System (LV) | 3.3 kg | 10% | 3.6 kg |
| Launch Mass | | | 904 kg |
| **Launch Vehicle Evaluation** | | | |
| Taurus II Enhanced Capability C3 = 2.0 | | | 1240 kg |
| Throw Mass Margin | | | 336 kg |

## 3.6 System Mass Budget and Launch Vehicle

Table 2 shows the EASCO system mass budget and the launch vehicle evaluation. A contingency of ~29% on the payload mass results in the mass allocated to the science instruments is ~178 kg. All masses are based on current best estimate (CBE). Based on the standard practice spacecraft bus and structural design, the S/C dry mass is estimated to be ~559 kg. The major portion of S/C mass is for the mechanical system (247 kg) followed by the power system (99 kg). Others are communications (70 kg), propulsion assembly (45 kg), attitude control (42 kg), thermal (30 kg), and avionics (26 kg). About 30% contingency was allowed for all instruments. An estimated 17.5% was allowed for the high TRL S/C bus based on GSFC GOLD rules resulting in a mass allocation of 657 kg for the bus. The total mass for the EP-MH hybrid propulsion system is ~65 kg, resulting in the S/C wet mass of ~900 kg.

Including the mass of the separation system of the launch vehicle, the resulting launch mass of 904 kg fits well within the capability (1240 kg) of a Taurus II with an enhanced faring and provides a throw mass margin of 336 kg (37%). There are several other launch vehicles that can be used, but they are generally heavier. When there is definite observatory growth, one may consider other launch vehicle options.

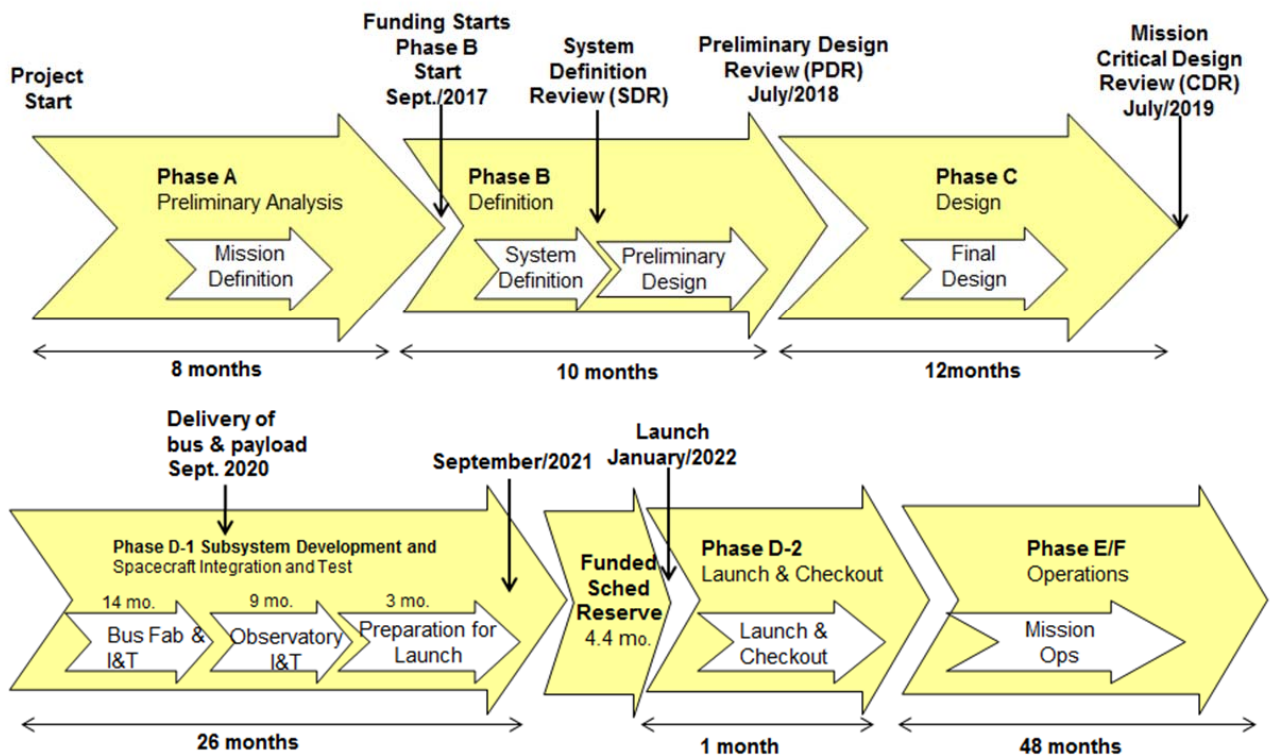

Figure 5. Project life cycle of the EASCO mission, showing the extents of various phases and the tasks to be completed.

### 3.7 EASCO Project Life Cycle

Figure 5 shows the project life cycle of the EASCO mission. From the start of Phase B, it takes about 48 months to launch, with an additional 4.5 months as funded schedule reserve. Assuming that the Phase B starts in September 2017, the L5 mission launch will be in January 2022. The critical design review (CDR) will be scheduled for July 2019, marking the end of phase C. Phase D-1 will last for 26 months, involving subsystem development and S/C Integration and Testing (I &T). It is assumed that the spacecraft bus will be built and tested out-of-house. Integration of S/C bus and the ten instruments and testing will also be done out-of-house and delivered in September 2020. Preparations for launch will start after observatory I & T and will last for ~3 months. The launch and check out will last for a month before starting the operations (Phase E/F). The prime mission will last for 48 months and the extended mission for additional 96 months.

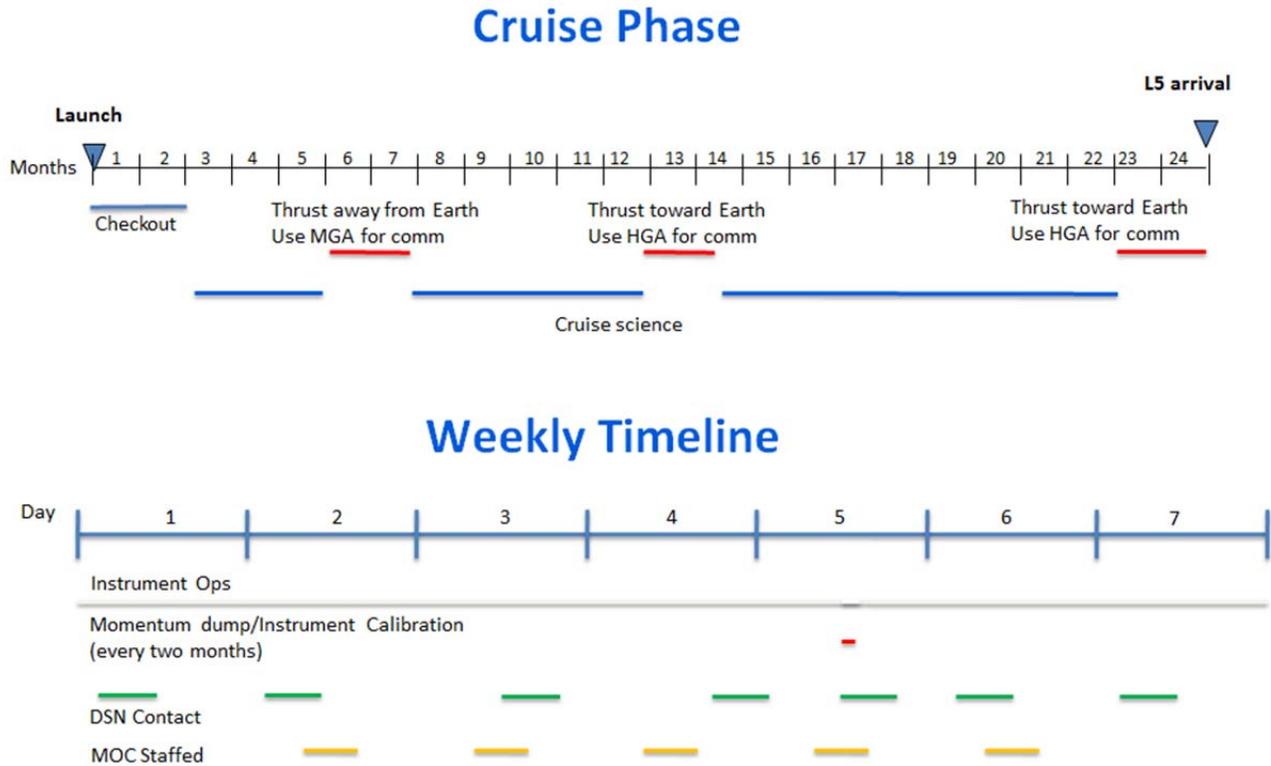

Figure 6. Cruise phase (top) and weekly operations time lines of the EASCO mission.

### 3.8 EASCO Operations

The Mission Operation Center (MOC) and the Science Operation Center (SOC) will be located at GSFC. The MOC will handle the following functions: mission planning and scheduling, orbit determination/control, network and contact scheduling, commanding, S/C monitor/control, real time health/safety processing, trending/analysis, instrument data handling, level 0 product processing, and level 0 data archive. The MOC implementation will use existing tools and software. There are no mission requirements that drive technology; technology required is readily available and operational today for several spacecraft.

The MOC will handle infrequent calibration rolls (MADI: offpoint and roll, WCOR: roll approximately once every two months), momentum dumps (approximately every two months, close instrument doors for the six remote sensing instruments), deployments (solar array, MAG boom, and LRT antennas), orbital maneuvers (three solar electrical propulsion thrust periods with a total period of 144 days). The thrust periods occur in the cruise phase (duty cycle is ~20%) and last for weeks. LRT antennas and Mag boom will not be deployed until after arrival at L5. When thruster is off, EASCO can take science data by opening the reclosable doors. Figure 6 (top) shows the cruise phase time line indicating the thrust periods and possible periods of cruise science.

DSN will be used to cover all the critical events of the mission: separation from launch vehicle, attitude acquisition, solar array deployment, and propulsion system tests. If the separation is not in the view of the ground station, a portable ground station will be used. The mission operation plan includes five elements: (i) Nominal Sequence Planning and Commanding: receive instrument commands from SOC five days per week (see Fig. 6 bottom); uplink command sequences every week day from the MOC. (ii) Operations staffing: 8 hours per day, five days per week operations by MOC staff; autonomous monitoring when unstaffed; designated operations team members will be alerted in the event of a problem or opportunity. (iii) Operations Training: operations team will participates in spacecraft integration and testing; also performs mission simulations prior to launch to verify readiness. (v) Operations Center Development: Reuses existing facility and software.

### 3.9 Other Considerations

EASCO will use a star tracker (roll knowledge) and a guide telescope (pointing accuracy) similar to the ones on STEREO, given the expected momentum dump every 2 months. A detailed analysis of jitter sources (booms for science instruments, gimbaled HGA, gimbaled Xenon thrusters, reaction wheels) needs to be carried out. The EASCO avionics includes (i) Integrated Avionics Unit: uses typical command and data handling cards; also handles data storage (50 GB solid state recorder), attitude Control, and S/C power/battery management. (ii) Redundancy Management Unit: manages primary and redundant power interfaces to Battery/Solar Array. (iii) Gimbal Control Electronics: controls dual axis gimbals for propulsion and antennas. The flight hardware/software is based on proven in-house or commercial off the shelf (COTS) system and no credible technical risk has been identified. EASCO thermal design meets all temperature requirements and conductivity requirements (Thermal Blanket grounding and shielded cabling); provides instrument suite thermal control requirements; provides an anti-sun side for radiating instruments and s/c loads. The radiation analysis showed that the dose is dominated by the protons and ions from the solar energetic particle events. The study recommended that dose be reduced to 15 krad (Si) if feasible. This may not be possible for all subsystems, but serves as a good guideline. The recommended dose level can be achieved with a 4.3 mm aluminum shielding. Designs are validated with appropriate reliability analyses – Fault Tree Analysis, FMEA, Parts Stress Analysis, Probabilistic Risk Analysis, and Worst Case Analysis. The design meets requirements for a Class B Mission with a probability of success (Ps) = 0.800 at 6 years. MDI and ICE (as of 3/14/2011) instruments have a very high likelihood of failure beyond 9 years (if single string piezoelectric mirror stabilization/kinematic mount system is used).

## 4. SUMMARY AND CONCLUSIONS

EASCO is a single observatory carrying 10 science instruments, seven of them remote-sensing and three of them in-situ. The spacecraft is three-axis stabilized and orbits in an Earth-like heliocentric orbit, behind Earth, at the Earth-Sun L5. EASCO is like a better SOHO with full in situ instruments and a better view of the Earth-directed CMEs. It is also like a better STEREO because it is stationed at L5 for long-term observations and carries a magnetograph. The Sun-Earth L5 is the next logical location for observing large-scale solar disturbances in the heliosphere and especially the disturbances that affect Earth. The mission targets the solar maximum in the year 2025, starting the observations one year before and 3 years after the maximum. The extended mission will continue to the following maximum 11 years later. EASCO will observe CMEs and CIRs from their source regions and as they pass through interplanetary space until they directly affect the near-Earth space environment. The L5 view provides a broadside view of the CMEs that produce solar energetic particles at Earth and geomagnetic storms. The CIRs will be observed in situ about 4 days ahead of time before they arrive at Earth.

The mission design laboratory study found that the EASCO Mission is considered very achievable with no new technology required. The key to the simple yet very flexible concept is the use of existing, flight proven, electric propulsion system hardware. Unlike most science missions, EASCO will use solar electric propulsion, which provides enormous flexibility in payload mass and launch vehicle selection. It was found that a medium launch vehicle is adequate (e.g., Taurus II with enhanced faring). The system design using electric propulsion is an efficient, elegant solution to meeting the mission requirements. All other subsystems are well within standard capabilities and borrow directly from the successful STEREO mission and from the SOHO and SDO missions.

ACKNOWLEDGMENTS

We thank all the GSFC Mission Design Laboratory engineers led by J. E. Oberright and M. Steiner who performed the study in collaboration with the EASCO team: J. Garrick, B. Lorenz, R. Vento, B. Beaman, F. Vaughn, H. Safren, L. Phillips, D. Peters, S. Riall, S. Tompkins, D. Batchelor, J. Panek, K. Brown, D. Steinfeld, J. Sturgis, D. Olney, P. Kalmanson, and R. Mills. Work supported by NASA/GSFC Internal Research and Development (IRAD) funds.